# Native Mass Spectrometry and Nucleic Acid G-quadruplex Biophysics - Advancing Hand in Hand

Valérie Gabelica

Univ. Bordeaux, CNRS, INSERM, ARNA, UMR 5320, U1212, IECB, F-33600 Pessac, France

v.gabelica@iecb.u-bordeaux.fr

**Conspectus**

While studying nucleic acids to reveal the weak interactions responsible for their three-dimensional structure and for their interactions with drugs, we also contributed to the field of biomolecular mass spectrometry, both in terms of fundamental understanding and with new methodological developments. A first goal was to develop mass spectrometry approaches to detect non-covalent interactions between antitumor drugs and their DNA target. Twenty years ago, our attention turned towards specific DNA structures such as the G-quadruplex (a structure formed by guanine-rich strands). Mass spectrometry allows to discern which molecules interact with one another by measuring the masses of the complexes, and quantity the affinities by measuring their abundance. The most important findings came from unexpected masses. For example, we showed the formation of higher- or lower-order structures by G-quadruplexes used in traditional biophysical assays. We can also derive complete thermodynamic and kinetic description of G-quadruplex folding pathways by measuring cation binding, one at a time. Getting quantitative information required accounting for nonspecific adduct formation and for the response factors of the different molecular forms. But with these caveats in mind, the approach is mature enough for routine biophysical characterization of nucleic acids. A second goal is to obtain more detailed structural information on each of the complexes separated by the mass spectrometer. One such approaches is ion mobility spectrometry, and even today the challenge lies in the structural interpretation of the measurements. We showed that, although structures such as G-quadruplexes are well-preserved in the MS conditions, double helices actually get more compact in the gas phase. These major rearrangements forced us to challenge comfortable assumptions. Further work is still needed to generalize how to deduce structures in solution from ion mobility spectrometry data, and in particular how to account for the electrospray charging mechanisms and for ion internal energy effects. These studies also called for complementary approaches to ion mobility. Recently, we applied isotope exchange labeling mass spectrometry to characterize nucleic acid structures for the first time, and we reported the first ever circular dichroism ion spectroscopy measurement on mass-selected trapped ions. Circular dichroism plays a key role in assigning the stacking topology, and our new method now opens the door to characterizing a wide variety of chiral molecule by mass spectrometry. In summary, advanced mass spectrometry approaches to characterize gas-phase structures work well for G-quadruplexes because they are stiffened by inner cations. The next objective will be to generalize these methodologies to a wider range of nucleic acid structures.

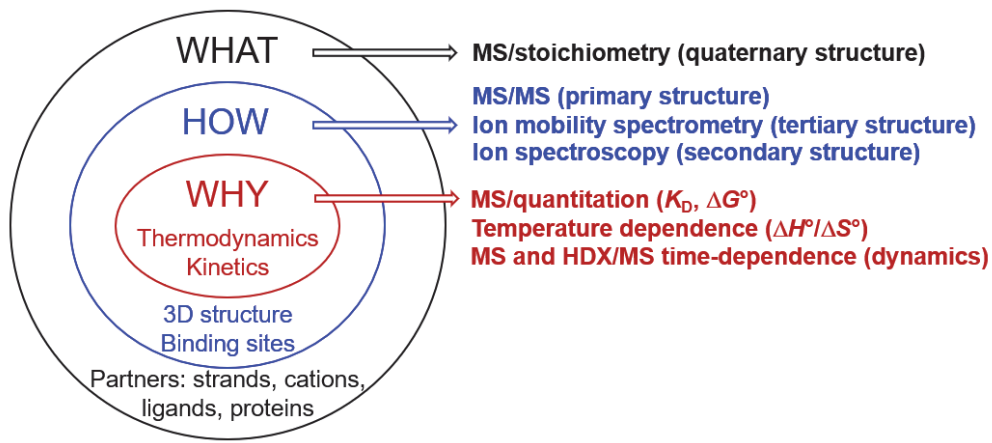

**Key references**

Marchand, A.; Gabelica, V. Folding and misfolding pathways of G-quadruplex DNA. *Nucleic Acids Res.* **2016**, *44*, 10999-11012.[1] *Here we used mass spectrometry to study the thermodynamics and kinetics of cation binding upon DNA G-quadruplex structure formation. We revealed that the most stable structures form slowly, because of the formation of competing misfolded intermediates.*

Marchand, A.; Rosu, F.; Zenobi, R.; Gabelica, V. Thermal Denaturation of DNA G-Quadruplexes and Their Complexes with Ligands: Thermodynamic Analysis of the Multiple States Revealed by Mass Spectrometry. *J. Am. Chem. Soc.* **2018**, *140*, 12553-12565.[2] *Here we investigated the temperature-dependence of complex equilibria using mass spectrometry. This led us to reveal the thermodynamics underpinnings of G-quadruplex formation and ligand binding, and in particular the entropic contributions, which cannot be grasped by structural biology alone.*

Porrini, M.; Rosu, F.; Rabin, C.; Darre, L.; Gomez, H.; Orozco, M.; Gabelica, V. Compaction of Duplex Nucleic Acids upon Native Electrospray Mass Spectrometry. *ACS Cent. Sci.* **2017**, *3*, 454-461.[3] *This paper challenged the widely accepted assumption according to which the Watson-Crick double helix shape is preserved inside a mass spectrometer. Instead, we show that more compact structures are formed. Our "negative results" explained many unpublished observations.*

Daly, S.; Rosu, F.; Gabelica, V. Mass-Resolved Electronic Circular Dichroism Ion Spectroscopy. *Science* **2020**, *368*, 1465-1468.[4] *We successfully transposed circular dichroism spectroscopy to measure the helicity of gas-phase ions. While applied here to DNA G-quadruplexes, this opens the door to taking advantage of the mass separation and circularly polarized light to characterize chiral molecules.*

**Introduction**

In the early 1990s, electrospray ionization (ESI) revolutionized biological mass spectrometry by enabling one to ionize large intact molecules,[5,6] and even preserve intact non-covalent complexes of proteins[7] or nucleic acids.[8-10] At the time, nucleic acids biophysics focused on understanding the driving forces of double-stranded DNA interaction with ligands (minor groove binders, intercalators, platinum complexes or cross-linking agents), which were mainly investigated for chemotherapy.[11-13] Hairpin-based regions in RNA were also hot target for antibiotics development.[14,15] More generally, the goal of nucleic acid ligand development is to hamper the division of specific cells, or to regulate gene expression. In the early 2000s, non-canonical DNA structures (hairpins, G-quadruplexes, i-motifs) attracted attention given their links to several diseases.[16] This is when we entered the field, with the motto "What can mass spectrometry do for nucleic acids biophysics?".

Biophysical studies ultimately aim at understanding *why* specific structures or complexes are formed. The "*why*" question can be reformulated as: "how thermodynamically and kinetically attainable is a given state, compared to all other possibilities?". The quantitative elements of answer are thus equilibrium constants, free energy, enthalpy and entropy differences between states, rates and rate constants.[17] Mass spectrometry characterization starts with the "*what*" question: what molecules interact with one another, in what proportions. Compared to other biophysics techniques,[18] mass spectrometry has the unrivaled ability to assign stoichiometries in a model-free manner. However, a mass measurement is blind to the three-dimensional arrangement. An active research area in mass spectrometry aims at developing methodologies (from concepts to instruments) to gather more information on *how* the complexes form, i.e., the pathways towards the final structure(s).

Most groups working on fundamental advances in mass spectrometry of non-covalent complexes are doing so on protein complexes. The need to verify the transposability of these approaches to nucleic acids forced us to keep an open mind, sometimes challenge pre-conceived ideas, and even develop new methods inspired by nucleic acids biophysics. In parallel, the problem of G-quadruplex folding turned out to be a formidable challenge for biophysics. G-quadruplex structures are formed when guanines assemble into G-quartets (Figure 1A) that stack onto one another thanks to cations. But predicting the folding is complicated, for two main reasons. First, some sequences can form multiple G-quadruplex topologies (in terms of strand stoichiometry, of which guanines form G-quartets together, strand orientation, geometry of the connecting loops). Second, many factors influence the topology: the oligonucleotide sequence and, in particular, the loop length and composition, flanking sequences, the cation nature and concentration, the presence of co-solvents or co-solutes, temperature cycles and reaction time scales,…. To date, although rules can predict impossible topologies, for lack of systematic quantitative data on the relative contribution of all influencing factors, for many sequences, we are still far from being able to predict what topology will form and why.

Over the course of collaboration with the G-quadruplex community, we focused on particular aspects of the topology that can uniquely be revealed with mass spectrometry (strand/cation/ligand stoichiometry), and developed methods to study this quantitatively or to gain more structural information on each of the stoichiometries. As a result, in our hands, advances in nucleic acids biophysics have been constantly intertwined with advances in measurement sciences. Our group recently reviewed mass spectrometry of nucleic acid complexes.[19] This more personal account highlights key novel insights generated through our effort in bridging the fields of nucleic acids biophysics and fundamental mass spectrometry.

**MS reveals unexpected stoichiometries and structures**

*STRANDS*

Because forming a G-quadruplex structure requires the parallel or antiparallel alignment of four tracts of guanines, an intuitive rule states that sequences containing one G-tract form tetramolecular G-quadruplexes, sequences containing two G-tracts form bimolecular G-quadruplexes, and sequences containing four G-tracts form intramolecular G-quadruplexes (Figure 1B—D). Early work in native MS of G-quadruplexes focused on sequences obeying this rule, such as [dTGGGGT]$_4$,[20] [dTTGGGGGT]$_4$,[21] [dGGGGTTTTGGGG]$_2$,[20] or dGGGTTAGGGTTAGGGTTAGGG.[20] This allowed us to validate MS-derived stoichiometries in the eyes of the biophysics community.

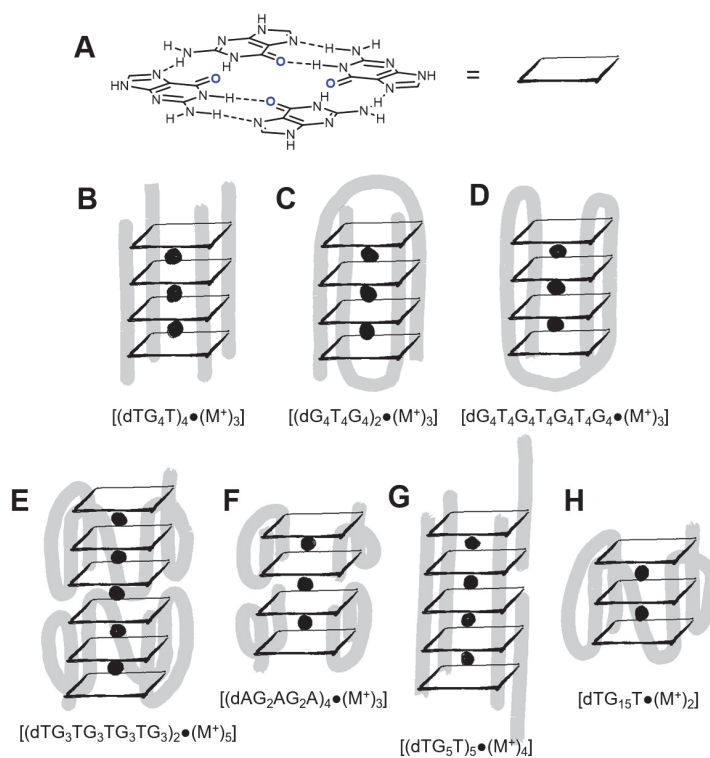

*Figure 1. Guanine quartets (A) and G-quadruplex structures: (B—D) structures wherein forming one G-quadruplex strictly requires four guanine tracts, and (E—H) unexpected strand and cation stoichiometries, contradicting this simplistic canonical rule, revealed by mass spectrometry: (E) cation-mediated dimer of a 4-repeat sequence,[22] (F) cation-mediated tetramer of a two-repeat sequence,[23] (G) pentamer transiently formed by a 1-repeat sequence,[24] (F) intramolecular G-quadruplex formed by a sequence with a single but long G-tract.[25]*

Later MS studies showed that this "predictive" rule was not always obeyed. The florilegia of exceptions to the rule includes: (1) sequences with four G-tracts forming multimeric structures (Figure 1E),[22,23,26] (2) sequences with two G-tracts forming tetramolecular structures (Figure 1F),[23,26,27] (3) sequences with one G-tract forming octamers besides or instead of tetramers,[28,29] (4) sequences with one G-tract forming pentameric intermediates (Figure 1G),[24] or (5) sequences with one G-tract forming tri-, bi-, and ultimately intramolecular structures (Figure 1H) as the length of the G-tract increases.[25] Sometimes the hunch for unusual stoichiometry was serendipitously observed by mass spectrometry, then confirmed by other biophysical methods. In other instances, the hunch for unusual stoichiometries came from gel electrophoresis, and the stoichiometry is unambiguously assigned using native MS.[28,30] A key point when cross-validating results obtained by different biophysical methods is to carry out part of the experiments in the same cation conditions.[22,23,28]

*CATIONS*

NH4OAc is the most widely used electrolyte in native mass spectrometry. The ionic radius of ammonium (1.54 Å),[31] closely resembles that of $K^+$ (1.51 Å).[32] Both cations coordinate in-between G-quartets, and thus forming $n$ quartets requires the specific binding of $n-1$ cations. When performing electrospray MS in ammonium, this preferred stoichiometry is most clearly observed for parallel-stranded G-quadruplexes, and was used to determine the number of effective quartets in sequences containing guanine mutations.[33] Also, based on the detection of an extra ammonium ion bound, we could infer which higher-order structures stem from the cation-mediated stacking of two G-quadruplex subunits.[26] However, a notable difficulty is that inter-quartet ammonium ions are lost (as $NH_3$) when the ion internal energy is increased, and this loss occurs at lower internal energy in non-parallel structures.[26,34] Hypotheses, still to be tested, are the proximity of loops for proton exchange when the cation exits by the top or bottom G-quartet, or the intrinsic stability of the G-quartet stacks (the octet coordination geometry can differ with the topology).

Eventually, the holy grail is to obtain MS information on G-quadruplexes in physiological conditions (in mammalian cells, typically 139 mM $K^+$, 12 mM $Na^+$ and 0.8 mM $Mg^{2+}$).[35] Whether sub-micrometer nanospray tips[36] will help achieve this goal must still be explored.[37] Our group mostly uses regular electrospray (2-4 µL/min) to save time, unless we are truly sample-limited. In 2014, we showed that by using trimethylammonium acetate (TMAA) instead of NH4OAc and up to 1 mM KCl, G-quadruplexes containing potassium (the physiologically relevant cation) could be formed.[38] We have been using these conditions for G-quadruplexes since then, especially for studies involving human telomeric sequences.[1,2,39-42] Recently, we studied a database of sequences of known structures in ~100 mM KCl and documented which structures are preserved in the MS-compatible conditions (1 mM KCl/100 mM TMAA).[42] Based on the $n$ quartets:$(n-1)$ cations rule, measuring the number of potassium ions specifically bound allowed us to reveal the formation of 2-quartet G-quadruplexes in sequences containing four tracts of three guanines, previously assumed to form 3-quartet structures.[1,39,43]

The binding of other cations than $NH_4^+$ or $K^+$ to G-quadruplexes has been studied as well, for example, the binding of alkaline earth metals.[43,44] Moreover, $Hg^{II}$ can bridge thymines in the loops[45] (note that some stoichiometries reported in the supporting information remain unexplained structurally), while $Ag^I$ induces the formation of silver-bound duplexes instead of G-quadruplexes with sequences containing one long G-tract.[46]

In summary, counting the strands and inner cations allows one to formulate structural hypotheses on the species formed. Because mass spectrometry enables an assumption-free determination of stoichiometry, it enables serendipitous discoveries, which are to our experience frequent with G-quadruplexes. There is already a treasure trove of seeing *what* is formed. Now, to delineate *why*, we need to quantify the energetic contributions involved.

**Advancing MS for quantitative biophysics**

SPECIFIC VS. NONSPECIFIC BINDING

When determining the stoichiometry of specifically bound cations or ligands, a key point is discerning the signals due to specific vs. nonspecific cation binding. A specific binding site is a specific binding location (and for ligands, a precise orientation of the ligand in a binding pocket). This translates into a higher binding affinity (lower equilibrium dissociation constant, $K_D$) than in alternative sites or orientations.

More difficult to grasp is the notion of *nonspecific* binding, because the term can have various meanings, one of them proper to mass spectrometry. On the one hand, a nonspecific complex can be a complex in alternative sites or orientations, existing in solution. For example, cations will be present in the counter-ionic atmosphere of negatively charged nucleic acids.[47] On the other hand, nonspecific binding can mean binding observed in the MS spectrum although it didn't exist in solution. This phenomenon occurs when droplets containing the analyte and excess binder evaporate, and increase the binder concentration before producing detectable complexes. The effect can be minimized by producing smaller droplets[37] or using lower binder concentrations, but depending on the affinity constant, the latter is not always possible. Alternatively, one can determine the fraction of the signal due to nonspecific binding by comparing the binding to the nucleic acid of interest and to a control sequence (thorough description can be found in the supplementary materials of earlier papers[1,39]). In that case, the exact definition of what is specific vs. nonspecific depends on the chosen control. For example, to determine the number $n$ of specific inter-quartet cations, we used control sequences with the same total length, same number of guanines, but unable to form G-quadruplexes in solution (see Figure 2). In such case, "nonspecific" binding means "non-G-quadruplex specific" binding, without presuming if it exists in solution. However, in our experience with up to 1 mM KCl, the amount of so-defined nonspecific $K^+$ binding depends on the ion charge state and on the source geometry, a strong hint that extra binding is linked to the electrospray process.

GETTING QUANTITATIVE: THE RESPONSE FACTORS

Next, one can obtain equilibrium constants for the formation of a specific complex from the relative mass spectral peak areas of free and bound stoichiometries. When mixing known amounts of target and binder, and based on the mass balance equation (the total amount of nucleic acid being distributed among several binding stoichiometries detected at equilibrium), one can determine the equilibrium concentration of each species if one knows their relative response factors, i.e. how relative peak areas in the mass spectra relate to relative concentrations in solution.

Although in early work we explored how to use relative intensities in dilution series to infer relative response factors,[48] in recent work we prefer using an internal standard to which all peak areas are compared.[49] Titration series with a fixed amount of target and increasing amounts of binder—or kinetic series with temporal evolution of the stoichiometries at fixed nucleic acid concentration—give a series of linear equations from which we can extract relative response factors. The lessons learned are that determining the response factors is essential to quantify species of different mass and size (for example, a tetramer → octamer reaction[28,50] or a G-quadruplex + complementary sequence → duplex formation).[49,51] In contrast, the response factors of free and cation-bound or small molecule ligand-bound nucleic acids differ by a factor of maximum 2 (at a given charge state, for experiments carried out in 100-150 mM ammonium or trimethylammonium acetate). Given that in thermodynamics, what matters is the order of magnitude of the equilibrium constant ($\Delta G° = -RT \ln K$), such difference in response factor is no more of a concern with native MS than with any other biophysical method.

Thus, although we study the binding of cations to multiply charged anions (in solution, one negative charge per phosphate group), the charge state distribution or the response relative response at a given charge state does not change much upon cation binding. At the ionic strength we typically use, volatile counterions neutralize about 80% of the charges. The remaining negative charges can be distributed in multiple ways, and the charge density is low enough that globular structures can accommodate extra negative charges on phosphate groups when additional cations are bound.

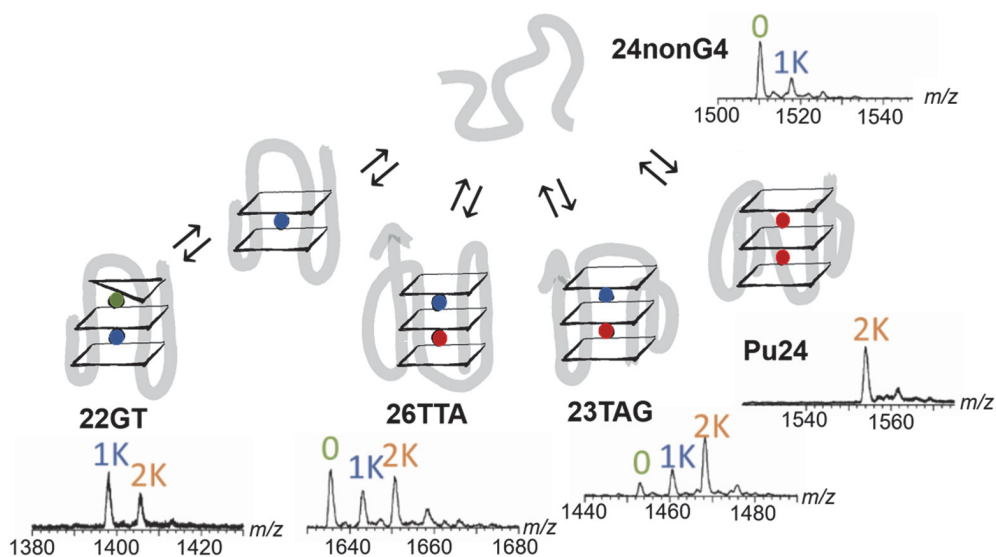

*Figure 2. Scheme of the competing folding pathways of 3-quartet intramolecular G-quadruplexes in potassium: Antiparallel 2-quartets structures form fastest, followed by hybrid-2, hybrid-1 and finally parallel structures. The zooms of the mass spectra show the potassium adduct distribution on the 5- charge state, recorded for 10 µM oligonucleotide and 500 µM KCl for a non-G4-forming sequence (24nonG4 = dTG$_3$ATGCGACAGAGAG$_2$ACG$_3$A), a 2-quartet forming sequence (22GT = dGGG(TTAGGG)$_3$T), a hybrid-2 forming sequence (26TTA = dTTAGGG(TTAGGG)$_3$TT), a hybrid-1 forming sequence that also forms a 2-quartet structure (23TAG = TAGGG(TTAGGG)$_3$) and a parallel G-quadruplex (Pu24 = dTGAG$_3$TG$_4$AG$_3$TG$_4$A$_2$G$_2$). The K$^+$ distribution on 24nonG4 corresponds to the non-G4-specific adducts distribution, and the K$^+$ distribution on the other sequences clearly shows the stability ranking of 3-quartet structures, with Pu24 being the most stable. The color code for the inner cations indicates the type of G-quartet stack they are surrounded with: red for homo-stacking, blue for heterostacking, and green for quartet/triplet stacking. Data taken from reference 1.*

# EXPLORING THERMODYNAMICS AND KINETICS OF G-QUADRUPLEXES WITH MS

The advantage of MS is the ability to obtain individual equilibrium constant for each ligand addition. and the ratio between consecutive equilibrium dissociation constants ($K_{D1}$, $K_{D2}$,...) informs on the binding cooperativity.[52] The $\Delta G°$ is directly deduced from the equilibrium constants, and with temperature-controlled nano-electrospray sources, the analysis of the temperature dependence of $\Delta G°$ provides the $\Delta H°$ of dissociation, and by difference, the $\Delta S°$. Combining the thermodynamic data gathered on K$^+$ binding to G-quadruplexes,[1,2] we found a $\Delta G°$ around $-7.5$ kcal/mol (all values ± 1 kcal/mol) for K$^+$ binding in-between same-polarity G-quartet stacks (red in Figure 2), $-5$ kcal/mol for opposite-polarity G-quartet stacks (blue in Figure 2), and $-4$ kcal/mol for quartet-triplet stacks (green in Figure 2). We also found that the quartet-triplet binding is entropically favorable, and thus more stable at higher temperature than the purely enthalpy-stabilized G-quartet stacks.

Much remains to be explored in terms of thermodynamics, however. Potassium binding to partial 6-ring G-quartet stacks (found in regular antiparallel 3-quartet topologies) has not yet been measured. Nor has cation binding thermodynamics to other special G-stacks at inter-subunit interfaces or in the left-handed Z-G4. Comparison of potassium with sodium, ammonium or other ions might also reveal interesting differences, which could contribute to shed light on the cation-dependence of G-quadruplex topology.

More detailed kinetics studies are necessary. Fitting cation binding kinetics provides additional $k_{on}$ and $k_{off}$ values, from which hidden $K_D$ values of sub-species (for example, kinetically distinct topologies of the same stoichiometry) can be extracted.[1] In a recent study, we monitored $^{41}$K/$^{39}$K exchange kinetics to determine the potassium residence time in the outer and inner G-stacks of the left-handed Z-G4.[53] Such time-resolved experiments can also be carried out in a temperature-dependent manner,[54] with the potential to provide potential energy landscapes with unprecedented level of detail.

Another aspect is the structural dynamics at a given temperature, for example, transient opening/closing of G-quartets in each topology. In G-quadruplexes, such events are monitored by the H/D exchange (HDX) of imino protons, typically by NMR.[55] We recently combined HDX with native MS, with a protocol involving complete deuterium labeling followed by time-resolved back-exchange.[56] The degree of labeling depends on the solution structure. Some deuteron lifetimes in G-quadruplexes can extend up to hours or days, while in duplexes the lifetimes are in the low-minute timescale and their kinetic study requires mixing devices. Conveniently, there is no memory of the electrospray charging process on the HDX extent, as all phosphates are neutralized by $^1$H protons. Our perspective is to obtain global thermodynamic and dynamic information on various G-quadruplex topologies, acquire local information by measuring the exchange kinetics at individual guanines, and explore how ligand or protein binding affects the dynamics.

**Nucleic acid structures in the gas phase**

Above, we showed the power of mass spectrometry to separate stoichiometries and to determine their associated thermodynamic and kinetic quantities. Yet each stoichiometric ensemble can still consist of a mixture of topologies. Ion mobility spectrometry coupled with mass spectrometry allows to separate each *m/z* ensemble according to shape: the ions are separated according to the friction caused by collisions with a gas, and in the low-field limit, the more compact conformations undergo less friction and travel faster than extended conformations.[57-59] We exploited this *separation capability* in our study of G-quadruplex formation kinetics: the kinetics of potassium binding suggested three kinetically distinct 2-$K^+$ complexes, and three peaks were indeed observed with ion mobility spectrometry, their relative abundance changing over time.[1] Mobility separation also helps to distinguish parallel and antiparallel forms of the bimolecular G-quadruplexes formed by telomeric repeats: the peak of the parallel form is narrower and the conformation is less compact than the antiparallel form.[34] Ion mobility spectrometry thus allows further parsing the different supramolecular ensembles formed in solution.

Beyond its separation capabilities, the holy grail in ion mobility spectrometry is to *attribute structures to each peak*. This assignment is based on the collision cross section (CCS), a quantity with the dimensions of a surface derived from the mobility measurement in a given gas and temperature, and which can be calculated for candidate structures (e.g., atomic models).[60] The first applications to duplex and G-quadruplex DNA were pioneered by the Bowers group, who concluded that nucleic acid structures were generally well preserved from solution to the gas phase.[61-66] These early experiments were carried out on a low-resolution drift tube ion mobility spectrometer, but later experiments on higher-resolution instruments revealed that some molecular systems display ion mobility peaks as narrow as a single-collision cross section peak shape (parallel G-quadruplexes[3,34,67] including the left-handed ones,[53] parallel silver-stabilized G-duplexes,[46] polythymines[67]), while others showed broad CCS distributions at all charge states (DNA and RNA Watson-Crick duplexes,[3,46] RNA hairpins and their complexes[68]), as exemplified in Figure 3 for 24-nt DNA duplex (the Dickerson-Drew dodecamer[69,70]) and parallel tetramolecular G-quadruplex ([dTGGGGT)$_4$●(NH$_4^+$)$_3$].[71]

Narrow peaks indicate either one rigid structure, or an ensemble of structures inter-converting in the gas phase on the millisecond time scale of the mobility separation. In G-quadruplexes, the core of G-quartet stacks likely remains rigid (as long as the inter-quartet cations are preserved), while terminal base and loop rearrangements are less hampered. Narrow peaks for sequences containing loops suggest they are locked in a specific conformation. Narrow peaks are observed with single-nucleotide loops of intramolecular parallel structures,[42,53] or for telomeric DNA mutants harboring TA base pairs in solution (e.g., 24TTG[42,72] or 5YEY[42]), which we deduce are preserved in the gas phase. In contrast, G-quadruplexes with longer loops or termini that are not pre-locked by specific base pairs end up forming broad peaks in ion mobility spectrometry.[42] This is due to the multiplicity of ways these parts can rearrange and form hydrogen bonds that, once locked in the gas phase, result in an ensemble of non-interconverting conformations. These multiple possibilities of H-bond networks can potentially reflect structural heterogeneity in solution. However, the phenomenon can also be purely due to the electrospray process, as we demonstrated for canonical duplexes.

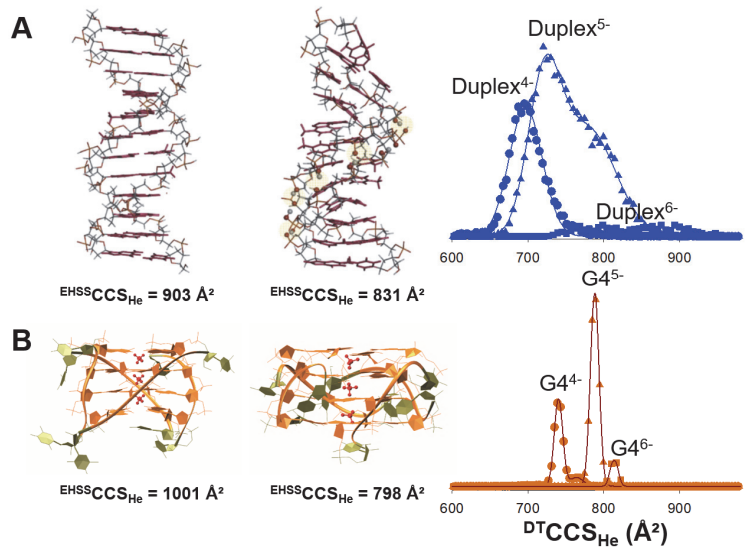

*Figure 3. Collision cross section distribution reconstructed from helium drift tube ion mobility measurements for the duplex [dCGCGAATTCGCG]$_2$ (top) and for the G-quadruplex [(dTGGGGT)$_4$•(NH$_4^+$)$_3$]. Far left: starting canonical B-helix duplex structure and G-quadruplex X-ray crystal structure. Middle: result of gas-phase modeling (A) at the semi-empirical level PM7 for the duplex$^{5-}$, showing the formation of P−O$^-$···H$^+$···$^-$O−P bonds across the minor groove, and (B) with the parmbsc0 force field for the G-quadruplex, showing good agreement with a preserved G-quadruplex core structure. Note that PM7 modeling matches as well. Data taken from references 3, 4 and 70.*

Baker and Bowers showed that for GC-rich duplexes, the collision cross sections were compatible with a preserved B-helix for DNA duplexes longer than ~20 base pairs and with ~1 charge per two base pairs.[62,65] Shorter duplexes with that charge density compacted to the level of A-helices.[65] However, when we measured the collision cross sections of DNA and RNA duplexes with lower charging levels, typical of electrospray ionization in 100-150 mM ammonium acetate, even smaller collision cross sections were found.[3] In Watson-Crick duplexes, the entire backbone can rearrange when going from the solution to the gas phase. Molecular modeling (whether force field, semi-empirical or DFT) reveals that charged (P−O$^-$) and neutralized (P−OH) phosphate groups can come together to share protons, forming P−O$^-$

···H$^+$···$^-$O−P networks.[3] In single strands, base-phosphate hydrogen bonds can also form.[73,74] This means that hydrogen bond interactions that did not exist in solution can form in the gas phase. But for molecules which backbones can rearrange in the gas phase (or in the final desolvation/declustering stages of the electrospray process) to make non-specific interactions, how to infer a solution conformation (or conformational ensemble) from the collision cross section distribution? This is where we reach the current limits of structural ion mobility spectrometry: control experiments hint at the preservation of some memory of the solution conformation, yet the conformation is affected by the measurement, for example by the electrospray charging process,[75] ion internal energy and time spent in the gas phase. The charging process is critical for nucleic acid rearrangements, as the phosphate groups are positioned along the backbone, and partial neutralization provides protons with the ability to form non-native hydrogen bonds, and which could be mobile.[73]

In summary, fifteen years ago we started with a naïve view of nucleic acid structures being well-preserved from the solution to the gas phase, and ion mobility spectrometry as a simple way to assign plausible/incompatible solution structures based on the collision cross sections.[61,63] This works only for rigid structures, and some G-quadruplexes were part of them. Now, we realize that elaborating a generalized framework to interpret ion mobility mass spectrometry measurement for non-fully rigid structures remains challenging, because it requires understanding the entire analyte history, from bulk solution to detection, including their electrospray "charging" process, internal energy uptake and possible rearrangements. In particular, we will devote future work to tackling the effect of internal energy and Coulomb forces on the formation of non-native or disruption of native non-covalent bonds.

We also believe that raising the level of structural modeling will improve structural assignment. In our most recent work on rigid nucleic acid structures (little rearrangement between solution and gas phase), we obtained excellent match between calculated and experimental collision cross sections in helium, when the structures were optimized at the semi-empirical (PM7) level and then submitted to Born-Oppenheimer Molecular Dynamics.[4,46,53] But predictive modeling of rearrangements will be a much grander challenge.

**Advances in biomolecular ion spectroscopy**

Besides ion mobility spectrometry, we exploited rigid and well-preserved G-quadruplex structures to test other gas-phase structural probing methods, such as ion spectroscopy. The goal of ion spectroscopy is to monitor the ion interaction with electromagnetic radiation as a function of its wavelength. Most gas-phase ion spectroscopy is devoted to small molecular ions, where energy levels can be calculated with sufficiently high accuracy to interpret the measurements by matching the theoretical and calculated spectra. But as the structures we are interested in are too large for this classical workflow (typically > 20-nucleotide structures, more than 600 atoms), we instead have to interpret the data based on clever control experiments, for example by comparing a duplex or G-quadruplex structure with its single strands.

Infrared spectroscopy gives us information on vibrational energy levels. For non-covalent interactions, IR spectroscopy is particularly useful to probe hydrogen bond formation. We could show that, if we studied low charge states (typically, no more than 5 negatively charged phosphates out of 20–24) and that the ion transfer conditions minimized unnecessary internal energy uptake, hydrogen bonds were indeed preserved in gas-phase G-quadruplexes,[76] i-motifs,[77] and Watson-Crick duplexes with GC base pairs.[78] However, to date we found no clear evidence of gas-phase Watson-Crick AT base pairs.[78]

UV-vis spectroscopy provides information on the electronic energy levels, which can be influenced by the chromophore environment and thus by the three-dimensional structure as well. Our first paper showed a significant red-shift (smaller energy difference between ground and electronically excited states) for GC-duplexes and G-quadruplexes compared to their constitutive single strands.[79] However, more recent in-depth investigation showed that the charge state (more likely, the charge density) of the ion also influenced the shape of the gas-phase spectra.[74] Gas-phase ion spectra are indeed action spectra reconstructed from the efficiency of the action of the radiation on the mass spectra (e.g. fragmentation or charge loss), and their shape results from a convolution from the absorption efficiency (what traditional absorption spectra measure) and the action efficiency. When submitted to electronic excitation, multiply charged nucleic acid anions undergo electron detachment,[80-82] a process which physical underpinnings are still incompletely understood but seems to involve the purine (and especially, guanine) bases although chemical intuition would lead one to imagine that the negative charges are on the phosphate groups. The electron photo-detachment from nucleic acids deserves further study. Similar phenomena have been observed in solution,[83] and gas-phase experiments thus offer opportunities to study a biologically relevant phenomenon in a controlled environment.

In summary, despite the intrinsic interest of electron photodetachment action spectroscopy, it turned out to be inappropriate for our original goal, i.e. determining gas-phase structures. In solution as well, the differences in UV absorption between different nucleic acid structures are subtle, and only difference spectra can reveal them.[84] Circular dichroism spectroscopy is another difference spectroscopy method, much more widely used to assign G-quadruplex topologies.[85-87] Indeed, positive or negative signals at 295, 260 and 240 nm provide obvious information on the stacking between the G-quartets. Thus, purely inspired by solution spectroscopy of G-quadruplexes, we decided to try measuring the electronic circular dichroism effect on gas-phase nucleic acid multiply charged anions.[4] We reconstructed the CD spectra (Figure 4) by monitoring the difference in electron photodetachment efficiency between left- and right-circularly polarized laser light, divided by the average electron photodetachment efficiency (ΔePD/ePD, left axis on Figure 4), as a function of the wavelength between 220 and 295 nm. For several model G-rich rigid structures, the solution-phase and gas-phase spectra were similar. We then applied CD ion spectroscopy to characterize the co-existing 1-$K^+$ and 2-$K^+$ topologies in telomeric G-quadruplexes. In future work, we will need to streamline the recording of the gas-phase CD spectra, and explore how to extend the measurement to other wavelengths, for example, to characterize proteins, amyloid aggregates, or other artificial constructs with stacked chromophores.

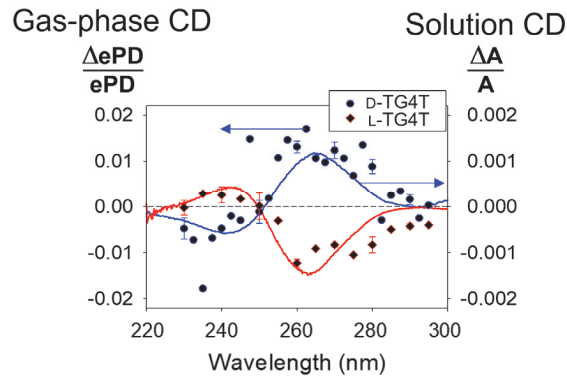

*Figure 4. The first gas-phase electronic circular dichroism spectrum of a biopolymer: comparison between solution CD (lines, right scale, relative difference of absorbance between left- and right-circularly polarized light) and gas-phase CD (symbols, left scale, relative difference of electron photodetachment yields between left- and right-circularly polarized light) for the right-handed parallel G-quadruplex formed from natural (L-sugars) $[(dTGGGGT)_4 \bullet (NH_4^+)_3]^{5-}$ (in blue) and its left-handed mirror image formed from the strand containing all D-sugars (in red). Data taken from reference 4.*

**Outlook**

This account illustrated how biophysical questions and methodological developments nourish each other. In the future, some important open questions in G-quadruplex biophysics will greatly benefit from mass spectrometry measurements. For example, quantifying the G-quadruplex folding thermodynamics and kinetics in terms of cation binding, one at a time, would enable one to predict the folding at various cation concentrations. The ultimate goal would be to link sequence, structure, and stability. In terms of structural analysis, based on the peak width in ion mobility spectrometry, one could screen for sequences able to lock the structures by specific hydrogen bonds. These structures would likely benefit from enthalpic stabilization, and are also more likely to be resolved in NMR. However, less stiff structures can still be rather stable thanks to favorable entropy contributions, which often forgotten in structure-based design because they are harder to picture. Flexible structures are also more difficult to characterize by NMR, but are amenable to MS. By combining temperature-variable native MS, ion mobility spectrometry and HDX-MS, we can imagine revealing the conformational entropy contribution of particular sequence regions, for a fuller picture of *why* particular structures are preferred.

Our studies also highlighted a number of unresolved important questions in mass spectrometry, which can benefit from using DNA, and G-quadruplexes in particular, as model systems. The most fundamental question is the relationship between the structure in the solution phase, in the gas phase, and in the step in-between—the transient "electrospray phase".[88] Nucleic acids offer unique opportunities because the solution-phase structure and stability can be fine-tuned by the sequence and modifications. This allows to test the influence of solution structure, solvent, additives and instrumental conditions on the extent of charging (or partial neutralization) and on gas-phase structures. From the point of view of structural characterization in the gas phase based on advanced MD methods, our work highlighted the importance of control experiments with mutated sequences to confirm the interpretations. Progress towards de novo interpretation of gas-phase experiments also compel further advances in gas-phase modeling, for example, polarizable and reactive force fields.

**Acknowledgements**

I would like to thank my team, and especially Frédéric Rosu for his comments on the present manuscript. The work described herein was made possible by the Inserm (ATIP-Avenir Grant no. R12086GS), the EU (FP7-PEOPLE-2012-CIG-333611, ERC-2013-CoG-616551-DNAFOLDIMS, H2020-MSCA-ITN-2014-ETN MetaRNA and H2020-MSCA-IF-2017-799695-CROWDASSAY) and the ANR (ANR-18-CE29-0013 POLYnESI).

**Author biography**

Valérie Gabelica was born in 1977 in Belgium, studied Chemistry and obtained her PhD in 2002 at the University of Liège (Belgium), with a thesis on mass spectrometry of non-covalent complexes under the mentorship of Prof. Edwin De Pauw. After a postdoc in Frankfurt (Germany) as Humboldt fellow in the group of Michael Karas, she rejoined the Mass Spectrometry Laboratory in Liège to study nucleic acid complexes by mass spectrometry, and she obtained a permanent position as FNRS research associate in October 2005. She joined the Institut Européen de Chimie et Biologie (IECB) in Bordeaux, France, in 2013 and became Research Director at the INSERM (French National Institute of Health and Medical Research). She is currently the director of IECB and associate editor of Analytical Chemistry. Her main research interests are fundamental aspects of mass spectrometry and its application to non-covalent complexes in general and nucleic acid complexes in particular, with research themes spanning from physical chemistry to biophysics, structural chemistry and biology.